\documentclass[10pt]{article}
\usepackage[numbers]{natbib}
\usepackage{graphicx}
\usepackage[a4paper, total={7in, 10in}]{geometry}
\usepackage{color, multirow,amsfonts,amsmath}
\usepackage{url}
\usepackage{rotating}
\begin{document}
\begin{center}
{\bf{Probabilities for asymmetric $p$-outside values}} - work in progress \\ Pavlina K. Jordanova

\bigskip

Faculty of Mathematics and Informatics, Konstantin Preslavsky University of Shumen, \\115 "Universitetska" str., 9712 Shumen, Bulgaria.\\ pavlina\_kj@abv.bg
\end{center}

\begin{abstract} In 2017-2020 Jordanova and co-authors investigate probabilities for $p$-outside values and determine them in many particular cases. They show that these probabilities are closely related to the concept for heavy tails. Tukey's boxplots are very popular and useful in practice. Analogously to the $\chi^2$-criterion, the relative frequencies of the events an observation to fall in different their parts, compared with the corresponding probabilities an observation of a fixed probability distribution to fall in the same parts, help the practitioners to find the accurate probability distribution of the observed random variable. These open the door to work with the distribution sensitive estimators which in many cases are more accurate, especially for small sample investigations. All these methods, however, suffer from the disadvantage that they use inter quantile range in a symmetric way. The concept for outside values should take into account the form of the distribution. Therefore, here, we  give possibility for more asymmetry in analysis of the tails of the distributions. We suggest new theoretical and empirical box-plots and characteristics of the tails of the distributions. These are theoretical asymmetric $p$-outside values functions. We partially investigate some of their properties and give some examples. It turns out that they  do not depend on the center and the scaling factor of the distribution. Therefore, they are very appropriate for comparison of the tails of the distribution, and later on, for estimation of the parameters, which govern the tail behaviour of the cumulative distribution function.  
\end{abstract}

\section{INTRODUCTION}

Along this work we consider a sample of independent identically distributed (i.i.d.) observations $X_1, ..., X_n$ on a random variable (r.v.) $X$, and corresponding increasing order statistics $X_{1:n} \leq X_{2:n} \leq \ldots X_{n:n}$. Denote the right continuous version of the cumulative distribution function (c.d.f.) of $X$ by $F_X(x) = \mathbb{P}(X \leq x)$, and its quantile function (generalized left-continuous inverse) by $F_X^\leftarrow(p) = \inf\{x \in \mathbb{R}: F_X(x) \geq p\}$, $p \in (0, 1]$. By assumption, $F_X^\leftarrow(0) = \sup\{x \in \mathbb{R}: F_X(x) = 0\}$, and $\sup \emptyset = -\infty$. The empirical $p$-quantiles can be defined in different ways. We restrict ourselves to the version, defined via the empirical c.d.f.
\begin{equation}\label{ECF}
F_n(x):= \left\{\begin{array}{ccc}
                                    0 & , & x < X_{1:n} \\
                                    \frac{i}{n} & , & x \in [X_{i:n};X_{i+1:n}) \\
                                    1 & , & x \geq X_{n:n} 
                                  \end{array}
\right., \quad x \in \mathbb{R}.
\end{equation}
Then, $F_n\left(X_{i:n}\right) = \frac{i}{n}$, $i = 1, 2, \ldots, n$. More precisely, we assume that  $F_n^\leftarrow(0) = X_{1:n}$, and for $p \in \left(\frac{i-1}{n}; \frac{i}{n}\right]$, i = 1, 2, \ldots, n, the empirical quantile function is
\begin{equation}\label{quantiles}
F_n^\leftarrow (p): = \inf\{x \in \mathbb{R}, F_n(x) \geq p\} = X_{\lceil np \rceil :n} = X_{i:n},
\end{equation}
 where $\lceil a \rceil$ means ceiling of $a$, i.e. the least integer greater than or equal to $a$. See for example \cite{hyndman1996sample,jordanova2020probabilities,serfling2009approximation}. It is implemented in function $quantile$, with parameter $Type = 1$ in the software R \cite{r2013r}. Then, $F_n^\leftarrow \left(\frac{i}{n}\right) = X_{i:n}$, $i = 1, 2, \ldots, n$.

Another good estimator could be found in \cite{parzen1979nonparametric}, p.18, and \cite{hyndman1996sample}. The authors use linear interpolation, and estimate the $p$-th quantile via $\hat{X}_{p, n} := X_{([(n+1)p],n)} + \left\{(n+1)p-[(n+1)p]\right\}\left\{X_{([(n+1)p]+1, n)} - X_{([(n+1)p], n)}\right\},$
where $[a]$ means the integer part of $a$, and $p \in \left[\frac{1}{n+1}, \frac{n}{n+1}\right]$. It also could be used for developing analogous to the following theory.   

In 1977-1978 Tukey et al. \cite{tukey1977exploratory,mcgill1978variations} defined box-and-whisker plots. The idea for this work originates from their works although the box-plots defined further on are not the same. In \cite{devore2015probability} there is a definition for mild and extreme outliers related to similar extremal box-plot. These works determine outliers as points isolated at different extent from the center of the distribution without taking into account the exact distribution of the observed r.v. Given a sample of independent observations on a r.v. of a fixed probability type, the authors of \cite{jordanova2017measuring} compute probabilities an observation to be mild or extreme outlier and suggest to classify the tails of linear probability types with respect to these probabilities. The last, compared with Devore's plots for extreme outliers, allows one to determine approximately the distribution of the observed r.v. with the most similar tail behaviour of the c.d.f. Later on, Jordanova and co-authors \cite{jordanova2018tails,jordanova2019probabilities,jordanova2020probabilities} generalize the results for arbitrary $p \in (0; 0,5]$. They investigate probabilities for $p$-outside values and determine them in many particular cases. All these results suffer from the disadvantage that they use the inter quartile range ($IQR$) in a symmetric way. More precisely, the middle of the interval between $p$-th and $1-p$-th quantile, coincides with the middle of the interval between the corresponding left and right $p$-fences. It is clear that this approach had been leading in the past, as far as, at that time, the outliers had been determined mainly with respect to the normal distribution.  It is well-known that the tail behaviour is not obligatory symmetric with respect to the center of the distribution. Here we give possibility for asymmetry in analysis of the tails of the distributions, and improve these results. We suggest new theoretical and empirical asymmetric $p$-box-plots and corresponding probabilities for asymmetric left and right $p$-outside values. Then, we partially investigate some of their properties and give some examples. These allows us to improve the existing classification of the distributions, with respect to their tail behaviour. Finally, we depict how these characteristics can help us to find the most appropriate classes of distributions for fitting the corresponding distribution of the observed r.v. and to estimate the parameters which govern the tail behaviour. The quality of the results about the suggested estimators is depicted via two examples. One of them considers a new estimator of the extremal index of Pareto distribution, and another one the index of regular variation of the tail of the c.d.f. in the linear Fr$\acute{e}$chet probability type.  

This work considers mainly right tails of distributions, in cases when there are positive probabilities for right $p$-outside values. There are many cases when these probabilities are zero. In that cases we do not need to estimate the parameter which governs the tail behaviour. Therefore, these cases are not the object of this study.   Analogous results for the left tails can be easily obtained when multiply the considered r.v. by $-1$. 

\section{DEFINITIONS AND PRELIMINARIES}
{\bf{Definition 1.}} Let $p \in (0; 0.5]$ be fixed. We call {\bf theoretical asymmetric right $p$-fence} the value of
\begin{eqnarray*}
R^A(F_X, p) &=& R^A(X,p) = F_X^\leftarrow(1 - p) + \frac{1-p}{p}(F_X^\leftarrow(1 - p) - F_X^\leftarrow(0.5)) \\
&=& \frac{1}{p}F_X^\leftarrow(1 - p) - \frac{1-p}{p}F_X^\leftarrow(0.5) = F_X^\leftarrow(0.5) + \frac{1}{p}(F_X^\leftarrow(1 - p) - F_X^\leftarrow(0.5)).
\end{eqnarray*}

The corresponding function of $p \in (0; 0.5]$ is called {\bf{theoretical asymmetric right fence function.}}

{\bf{Definition 2.}} Let $p \in (0; 0.5]$ be fixed. We call {\bf theoretical asymmetric left $p$-fence} the value of
\begin{eqnarray*}
L^A(F_X, p) &=& L^A(X,p) = F_X^\leftarrow(p) - \frac{1-p}{p}(F_X^\leftarrow(0.5) - F_X^\leftarrow(p)) \\
&=& \frac{1}{p}F_X^\leftarrow(p) - \frac{1-p}{p}F_X^\leftarrow(0.5) = F_X^\leftarrow(0.5) - \frac{1}{p}(F_X^\leftarrow(0.5) - F_X^\leftarrow(p)).
\end{eqnarray*}
\begin{figure}[h]
\centerline{\includegraphics [width=.4\textwidth]{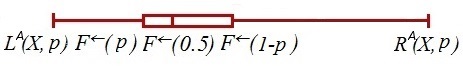}}
\caption{$p = 0.25$ \label{fig:AsymBoxPlotspOutliers}}
\end{figure}

The corresponding function of $p \in (0; 0.5]$ is called {\bf{theoretical asymmetric left fence function.}}

Figure \ref{fig:AsymBoxPlotspOutliers} depicts an example of the placement of asymmetric fences with respect to the center of the distribution in case when $p = 0.25$.

{\it Note:} 1. For any distribution $L^A\left(X, 0.5\right) = R^A\left(X, 0.5\right) = F_X^\leftarrow(0.5)$, and this case can be used for checking our results.

2. The theoretical asymmetric right fence function is non-increasing, and the left one is non-decreasing in $p \in (0; 0.5]$.

3. The fence functions preserve linear operations. More precisely, if $X$ is a r.v., and $a \in \mathbb{R}$ and $b > 0$ are constants, then for all $p \in (0; 0.5]$, 
\begin{equation}\label{affineinvariance}
R^A\left(\frac{X+a}{b}, p\right) = \frac{R^A(X, p) + a}{b} {\text{ and }} L^A\left(\frac{X+a}{b}, p\right) = \frac{L^A(X, p) + a}{b}. 
\end{equation}

4. If the scaling is negative, we switch from the left to the right tail, and conversely. For example, $L^A(X,p) = R^A(-X,p)$.

The following functions of $p \in (0; 0.5]$ are one and the same within the whole linear probability type. This allow ones, first to determine the most appropriate 
probability law, and then to estimate only the parameters which govern the tail behaviour of the considered distribution, by using parametric estimators.

{\bf{Definition 3.}} For any fixed $p \in (0; 0.5]$ we denote by
\begin{equation*}
p_{A,R,p}(X): = p_{A,R,p}(F_X):  = \mathbb{P}(X > R^A(F_X, p))
\end{equation*}
the probability of the event a r.v. $X$ to exceed the theoretical asymmetric right $p$-fence of the distribution of $X$, and call it {\bf probability for asymmetric right $p$-outside value}. The corresponding function $p_{A,R,p}(X)$ of $p \in (0; 0.5]$ is called {\bf{theoretical asymmetric right outside values function.}} 


{\bf{Definition 4.}} Let $p \in (0; 0.5]$ be fixed. We denote by
\begin{equation*}
p_{A,L,p}(X): = p_{A,L,p}(F_X):  = \mathbb{P}(X < L^A(F_X, p))
\end{equation*}
the probability of the event a r.v. $X$ to be less than the theoretical asymmetric left $p$-fence of the distribution of $X$, and call it {\bf probability for asymmetric left $p$-outside value}. The corresponding function $p_{A,L,p}(X)$ of $p \in (0; 0.5]$ is called {\bf{theoretical asymmetric left outside values function.}} 


The main, but easy to prove properties of these probabilities are that for any constants $a \in \mathbb{R}$ and $b > 0$, 
\begin{equation}\label{invariancepropery}
p_{A,L,p}(X) = p_{A,L,p}\left(\frac{X+a}{b}\right) {\text{ and }} p_{A,R,p}(X) = p_{A,R,p}\left(\frac{X+a}{b}\right).
\end{equation}
The last mean that they do not depend on the center and the scaling factor of the distribution. Therefore, they are very appropriate for comparison of the tails of the distributions, and estimation of the coefficients which govern the tail behaviour. Here again we have to clarify that if the scaling is negative, then we have to change the considerations and instead the left tail the results are for the right tail and conversely. The most important case is $p_{A,L,p}(-X) = p_{A,R,p}(X)$, because for any constant $b > 0$, $p_{A,L,p}(-bX) = p_{A,R,p}(X)$.

Another nice property is that these probabilities are invariant with respect to exceedance of the distribution. More precisely, for all  $p \in (0; 0.5]$, and $t \in \mathbb{R}$, such that $\mathbb{P}(X < t)> 0$,
$$p_{A,L,p}(X-t|X > t) = p_{A,L,p}(X|X > t), \quad p_{A,R,p}(X-t|X > t) = p_{A,R,p}(X|X > t).$$

{\it Note:} The asymmetric outside values depend on the distribution of the considered r.v. An observation can be asymmetric $p$-outside value with respect to (w.r.t.) a distribution, and for the same $p$, not to be asymmetric $p$-outside value w.r.t. another probability law. In this way one can chose that probability type which fits the data in the best way.

By plugin the estimators (\ref{ECF}) and (\ref{quantiles}) in the above four definitions we obtain the corresponding empirical counterparts.

{\bf{Definition 5.}} Let $p \in (0; 0.5]$ be fixed. The {\bf empirical asymmetric right $p$-fence} is
\begin{eqnarray*}
R_n^A(p) &=& F_n^\leftarrow(1 - p) + \frac{1-p}{p}(F_n^\leftarrow(1 - p) - F_n^\leftarrow(0.5)) \\
&=& \frac{1}{p}F_n^\leftarrow(1 - p) - \frac{1-p}{p}F_n^\leftarrow(0.5) = F_n^\leftarrow(0.5) + \frac{1}{p}(F_n^\leftarrow(1 - p) - F_n^\leftarrow(0.5)).
\end{eqnarray*}

The corresponding function of $p \in (0; 0.5]$ is called {\bf{empirical asymmetric right fence function.}}

{\bf{Definition 6.}} Let $p \in (0; 0.5]$ be fixed. The {\bf empirical asymmetric left $p$-fence} is
\begin{eqnarray*}
L_n^A(p) &=& F_n^\leftarrow(p) - \frac{1-p}{p}(F_n^\leftarrow(0.5) - F_n^\leftarrow(p)) \\
&=& \frac{1}{p}F_n^\leftarrow(p) - \frac{1-p}{p}F_n^\leftarrow(0.5) = F_n^\leftarrow(0.5) - \frac{1}{p}(F_n^\leftarrow(0.5) - F_n^\leftarrow(p)).
\end{eqnarray*}

The corresponding function of $p \in (0; 0.5]$ is called  {\bf{empirical asymmetric left fence function.}}

{\bf{Definition 7.}} Let $p \in (0; 0.5]$ be fixed. We call {\bf empirical asymmetric right(left) $p$-outside values} observations $X_i$, $i = 1, 2, \ldots, n$ which are bigger (smaller) than $R_n^A(p)$($L_n^A(p)$).

{\bf{Definition 8.}} The relative frequency of observations $X_i$, $i = 1, 2, \ldots, n$ which are bigger(smaller) than $R_n^A(p)$($L_n^A(p)$) is called {\bf{plug-in estimator of probability for asymmetric right(left) $p$-outside value.}} We denote them by $\hat{p}_{A,L}(p)$ ($\hat{p}_{A,R}(p)$).

Analogously to the results in Section 3.3, \cite{jordanova2020probabilities}, it can be shown that under relatively general conditions, the empirical asymmetric right(left) $p$-fences are asymptotically unbiased, asymptotically normal, weakly consistent and asymptotically efficient estimators for the corresponding theoretical asymmetric right(left) $p$-fences. Moreover, the last plug-in estimators are weakly consistent, asymptotically unbiased and asymptotically efficient estimators for corresponding probabilities for asymmetric $p$-outside values.

\section{PARTICULAR CASES}
Probabilities for extreme outliers (or "$0.25$-outside values"), for the most of the probability distributions considered in this section, are computed in \cite{jordanova2018tails}. Here we determine theoretical asymmetric fence functions, and theoretical asymmetric outside values functions.
We consider $p \in (0; 0.5]$, and due to property (\ref{invariancepropery}), without lost of generality, we chose the simplest parametric form of the considered linear probability type. 

\begin{figure}[h]
\begin{minipage}[t]{0.5\linewidth}
\centerline{%
    \includegraphics
    [width=.86\textwidth]{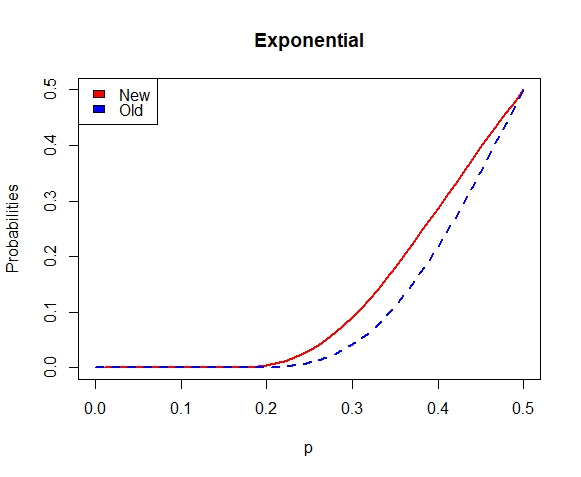}}
\end{minipage}
\begin{minipage}[t]{0.5\linewidth}
\centerline{%
    \includegraphics
[width=.86\textwidth]{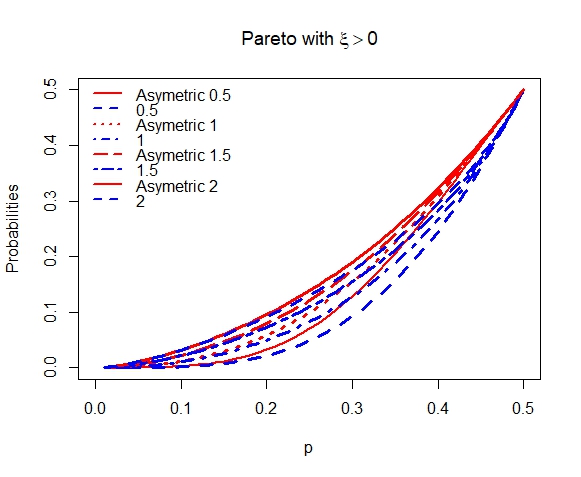}}
\end{minipage}
\caption{$p_{A,R,p}(X)$ (red line) and $p_{R,p}(X)$ (blue line) in the Exponential (left) and Pareto (right) cases. \label{fig:Exp}}
\end{figure}

{\bf Exponential $l$-type}. Let $X$ be a r.v. with c.d.f. $F_X(x) = 0$ for $x < 0$, and $F_X(x) = 1-e^{-x}$, otherwise. Then, $F_X^\leftarrow(p) = -\log(1-p)$, and $R^A(X,p) = -\frac{1}{p}\log(p)-\frac{1-p}{p}\log(2)$. In this case $L^A(X,p) = -\frac{1}{p}\log(1-p)-\frac{1-p}{p}\log(2)$. The plots of  $p_{A,R,p}(X) = p^{\frac{1}{p}}2^{\frac{1-p}{p}}$, and $p_{R,p}(X)$, (see \cite{jordanova2019general}) are presented in Figure \ref{fig:Exp}, left. For $p \in (0; p_0]$, where $p_0$ is the solution of $\log(1-p) + (1-p)\log(2) = 0$, i.e. $p_0 \approx 0,3588$, $p_{A,L,p}(X) = 0$. When $p \in (p_0; 0.5]$,  $p_{A,L,p}(X) = 1 - (1-p)^{\frac{1}{p}}2^{\frac{1-p}{p}}$. 
In particular, $p_{A, R, 0.25}(X) = \frac{1}{32} = 0.03125.$

{\bf Pareto $l$-type with positive shape parameter $\xi = \frac{1}{\alpha} > 0$}. In this case $X$ is a r.v. with c.d.f. $F_X(x) = 0$ for $x < 1$, and $F_X(x) = 1-x^{-\frac{1}{\xi}}$, otherwise. Then, $F_X^\leftarrow(p) = (1-p)^{-\xi}$, and $R^A(X,p) = p^{-1-\xi} - \frac{1-p}{p}2^\xi$. Now, $L^A(X,p) = \frac{1}{p(1-p)^{\xi}} - \frac{1-p}{p}2^\xi$. The plots of  $p_{A,R,p}(X) = \left(p^{-\xi-1} - \frac{1-p}{p}2^{\xi}\right)^{-\frac{1}{\xi}}$, and $p_{R,p}(X)$, (see \cite{jordanova2019general}) for different values of the shape parameter $\xi$ are presented in Figure \ref{fig:Exp}, right. We observe that the larger the parameter $\xi$, the greater these probabilities are. These correspond to the well-known result from extreme value theory that the greater the parameter $\xi$, the heavier the right tail of this distribution is. Analogously, for $p \in (0, p_0]$, where $p_0$ is the solution of $(1-p)^{\xi+1}2^\xi + p(1-p)^\xi = 1$,  $p_{A,L,p}(X) = 0$, and $p_{A,L,p}(X) = 1 - \left(\frac{1}{p(1-p)^{\xi}} - \frac{1-p}{p}2^\xi\right)^{-\frac{1}{\xi}}$, for $p \in (p_0; 0.5]$.

{\bf Pareto $l$-type with negative shape parameter $\xi = \frac{1}{\alpha} < 0$}. Let $X$ be a r.v. with c.d.f. $F_X(x) = 0$ for
$x < 0$,  $F_X(x) = 1-(1+\xi x)^{-\frac{1}{\xi}}$ for $x \in \left[0, -\frac{1}{\xi}\right)$, and $F_X(x) = 1$, otherwise.  Then,  $F_X^\leftarrow(p) = \frac{(1-p)^{-\xi}-1}{\xi}$, and  $R^A(X,p) = \frac{p^{-\xi}-1}{p\xi} - \frac{1-p}{p}\frac{2^\xi-1}{\xi} \geq 0$, for $p \in \left(0, \frac{1}{2}\right]$. 
In order to compute the probability $p_{A,R,p}(X)$ we need to know the solution of $R^A(X,p) < -\frac{1}{\xi}$.
\begin{eqnarray*}
  R^A(X,p) < -\frac{1}{\xi} & \iff & \frac{p^{-\xi}-1}{p\xi} - \frac{1-p}{p}\frac{2^\xi-1}{\xi}  < -\frac{1}{\xi} \iff -p^{-\xi} + 1 + (1-p)(2^\xi-1) < p  \\
   &\iff & (2p)^{-\xi} > 1-p \iff \xi > -\frac{\log_e(1-p)}{\log_e(2p)}.
\end{eqnarray*}

Therefore, if $\xi  > -\frac{\log_e(1-p)}{\log_e(2p)}$, and $p \in (0; 0.5)$,
$$p_{A,R,p}(X) = \left(1+\frac{p^{-\xi}-1}{p} - \frac{1-p}{p}(2^\xi-1)\right)^{-\frac{1}{\xi}} =  p\left(\frac{1}{p} - \frac{1-p}{p}(2p)^\xi\right)^{-\frac{1}{\xi}},$$  and $p_{A,R,p}(X) = 0$, otherwise. 

  Let us, now consider the smaller values than the median. In this case  $L^A(X,p) = \frac{(1-p)^{-\xi}-1}{p\xi} - \frac{1-p}{p}\frac{2^\xi-1}{\xi}$. Then, 
$$ L^A(X,p) > 0  \iff  (1-p)^{-\xi} - (1-p)2^\xi - p < 0 \iff 1 < (1-p)^\xi(2^\xi(1-p) + p).$$
 If we denote by $p_0$ the solution of $1 = (1-p)^\xi(2^\xi(1-p) +p)$, then  $p_{A,L,p}(X) = 0$, for $p \in (0; p_0]$, and 
  $$p_{A,L,p}(X) = 1-\left(1+\frac{(1-p)^{-\xi}-1}{p} - \frac{1-p}{p}(2^\xi-1)\right)^{-\frac{1}{\xi}} = 1 - (1-p)\left(1-\frac{1-p}{p}(2(1-p))^\xi\right)^{-\frac{1}{\xi}},$$ 
  for $p \in (p_0; 0.5]$.
  
  As far as this distribution has a bounded support we will not plot these probabilities.
  
  {\bf Fr$\acute{e}$chet $l$-type with parameter $\alpha > 0$.} Let us now consider a r.v $X$ with c.d.f. $F_X(x) = 0$ for $x < 0$, and $F_X(x) = \exp\{-x^{-\alpha}\}$, when $x \geq 0$. In this case the quantile function is $F_X^\leftarrow(p) = (-\log(p))^{-\frac{1}{\alpha}}$. Therefore,  definition 1 and definition 2 entail
  \begin{eqnarray*}
  L^A(X,p)  & = & \frac{1}{p}(-\log(p))^{-\frac{1}{\alpha}} - \frac{1-p}{p}(\log(2))^{-\frac{1}{\alpha}}\\
  R^A(X,p)  & = & \frac{1}{p}(-\log(1-p))^{-\frac{1}{\alpha}} - \frac{1-p}{p}(\log(2))^{-\frac{1}{\alpha}}.
  \end{eqnarray*}
 In this case $L^A(X,p) \geq 0 \iff  (-\log(p))^{-\frac{1}{\alpha}} \geq  (1-p)(\log(2))^{-\frac{1}{\alpha}} \iff  \left(\frac{\log(2)}{-\log(p)}\right)^{\frac{1}{\alpha}} \geq 1 - p  \iff \log\left(\frac{\log(2)}{-\log(p)}\right) \geq \alpha \log(1-p) \iff \alpha \geq \frac{\log\left(\frac{\log(2)}{-\log(p)}\right) }{\log(1-p)}.$
The last considerations mean that if we denote by $\alpha_0 : = \frac{\log\left(\frac{\log(2)}{-\log(p)}\right) }{\log(1-p)}$. Then, for $\alpha \in (0, \alpha_0]$, $p_{A,L,p}(X) = 0$, and for $\alpha  > \alpha_0$,
 $$p_{A,L,p}(X) = \exp\left\{-\left(\frac{1}{p}(-\log(p))^{-\frac{1}{\alpha}} - \frac{1-p}{p}(\log(2))^{-\frac{1}{\alpha}}\right)^{-\alpha}\right\}.$$

\begin{figure}[h]
\begin{minipage}[t]{0.5\linewidth}
\centerline{%
    \includegraphics
    [width=.86\textwidth]{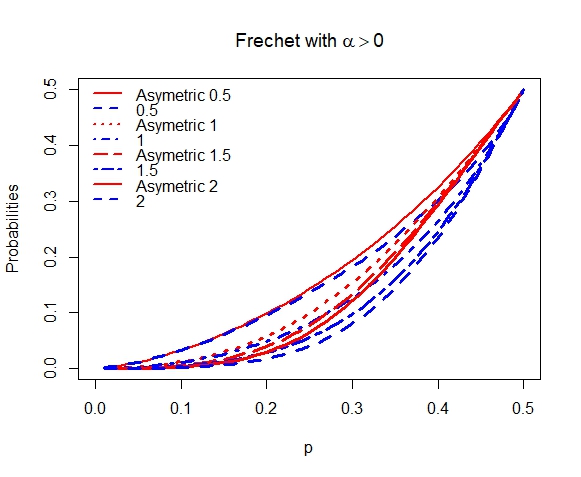}}
\end{minipage}
\begin{minipage}[t]{0.5\linewidth}
\centerline{%
    \includegraphics
[width=.86\textwidth]{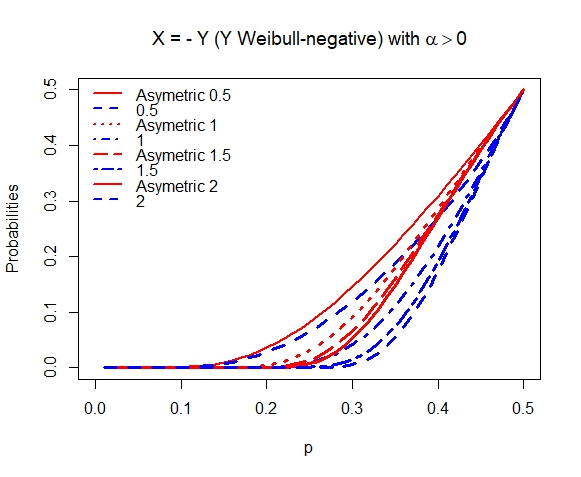}}
\end{minipage}
\caption{$p_{A,R,p}(X)$ (red line) and $p_{R,p}(X)$ (blue line) in the Fr$\acute{e}$chet (left) and Weibull (right) cases. \label{fig:FrechetAndWeibull}}
\end{figure}

Analogously, we obtain that for all $\alpha > 0$, and $p \in (0; 0.5]$, $R^A(X,p) \geq 0$. Therefore,
$$p_{A,R,p}(X) = 1 - \exp\left\{-\left(\frac{1}{p}(-\log(1-p))^{-\frac{1}{\alpha}} - \frac{1-p}{p}(\log(2))^{-\frac{1}{\alpha}}\right)^{-\alpha}\right\}.$$
In this case the plots of  $p_{A,R,p}(X)$, and $p_{R,p}(X)$, (the last notation is defined in \cite{jordanova2019general}) as functions of $p$, and for different values of the parameter $\alpha$ could be seen in Figure \ref{fig:FrechetAndWeibull}, left. We observe that the greater the parameter $\alpha$, the smaller these probabilities are. These correspond to the well-known result from extreme value theory that the smaller the parameter $\alpha$, the heavier the right tail of this distribution. 

 {\bf Weibull-negative $l$-type with parameter $\alpha > 0$, $\xi= -\frac{1}{\alpha} < 0$}. Together with the previous, and the next distribution, they form the class of Extreme value distributions, and they are all distributions that are max-stable with respect to linear transformations. Here, w.l.g. we use the most comfortable centering and scale parameters. More precisely, let $Y$ be a r.v. with c.d.f. $F_Y(x) = \exp\left\{-(-x)^{\alpha}\right\}$, when $x < 0$, and $F_Y(x) = 1$, otherwise. It is clear that in this case, the interesting tail is the left one. In order to compare it with the other right tails, instead of $Y$ we consider $X = -Y$. Therefore, $F_X(x) = 1 - \exp\left\{-x^{\alpha}\right\}$, when $x > 0$, and $F_X(x) = 0$, otherwise, and the quantile function is $F_X^\leftarrow(p) = (-\log(1-p))^{\frac{1}{\alpha}}$. Then,    \begin{eqnarray*}
  L^A(X,p)  & = & \frac{1}{p}(-\log(1-p))^{\frac{1}{\alpha}} - \frac{1-p}{p}(\log(2))^{\frac{1}{\alpha}}\\
  R^A(X,p)  & = & \frac{1}{p}(-\log(p))^{\frac{1}{\alpha}} - \frac{1-p}{p}(\log(2))^{\frac{1}{\alpha}}.
  \end{eqnarray*}
 Analogously to the previous case $ L^A(X,p) \geq 0 \iff -\log(1-p) \geq  (1-p)^\alpha \log(2) \iff   \frac{-\log(1-p)}{\log(2)} \geq  (1-p)^\alpha \iff  -\log\left(\frac{\log(2)}{-\log(1-p)}\right) \geq \alpha \log(1-p) \iff \alpha \geq \frac{\log\left(\frac{\log(2)}{-\log(1-p)}\right) }{-\log(1-p)}.$
Now, if $\alpha_0 : = \frac{\log\left(\frac{\log(2)}{-\log(1-p)}\right) }{-\log(1-p)}$, then for $\alpha \in (0, \alpha_0]$, $p_{A,L,p}(X) = 0$, and for $\alpha  > \alpha_0$,
 $$p_{A,L,p}(X) = 1-\exp\left\{-\left(\frac{1}{p}(-\log(1-p))^{\frac{1}{\alpha}} - \frac{1-p}{p}(\log(2))^{\frac{1}{\alpha}}\right)^\alpha\right\}.$$

Again, for all $\alpha > 0$, and $p \in (0; 0.5]$, $R^A(X,p) \geq 0$, and thus,
$$p_{A,R,p}(X) = \exp\left\{-\left(\frac{1}{p}(-\log(p))^{\frac{1}{\alpha}} - \frac{1-p}{p}(\log(2))^{\frac{1}{\alpha}}\right)^{-\alpha}\right\}.$$

These theoretical asymmetric right outside values functions, together with $p_{R,p}(X)$, defined in \cite{jordanova2019general}, for different values of the parameter $\alpha$ are plotted in Figure \ref{fig:FrechetAndWeibull}, right.  They show that the greater the parameter $\alpha$, the smaller these probabilities are. Moreover, when compare the left and the right panel of the same figure, for any fixed $p \in (0; 0.5]$ and $ \alpha > 0$, we observe that the  Fr$\acute{e}$chet $l$-type has heavier probabilities $p_{A,R,p}(X)$, than the probabilities $p_{A,R,p}(-Y)$ in the Negative-Weibull case. Again we have analogous results from extreme value theory for the heaviness of the right tails of these two distributions.

\begin{figure}[h]
\begin{minipage}[t]{0.5\linewidth}
\centerline{%
    \includegraphics
    [width=.86\textwidth]{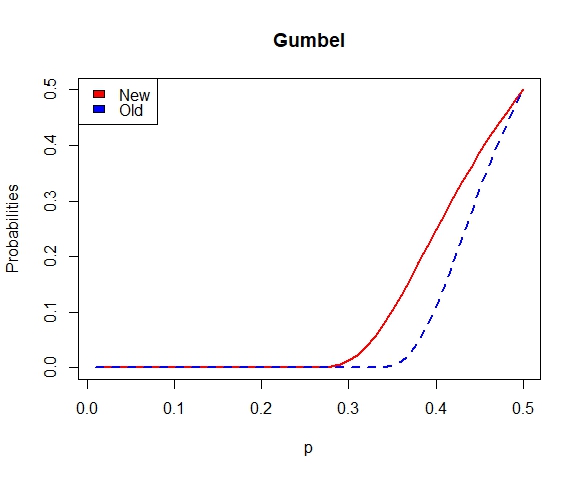}}
\end{minipage}
\begin{minipage}[t]{0.5\linewidth}
\centerline{%
    \includegraphics
[width=.86\textwidth]{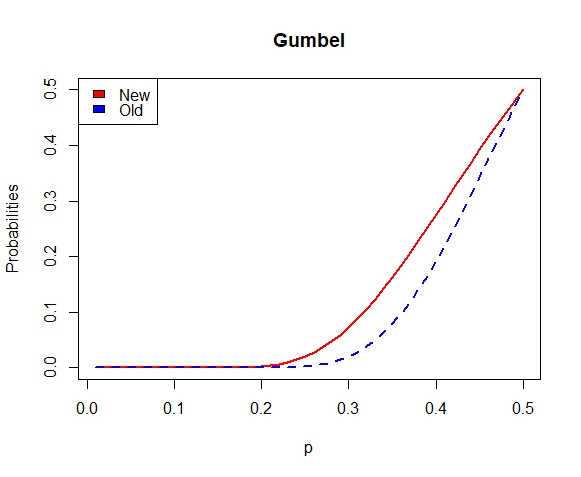}}
\end{minipage}
\caption{$p_{A,L,p}(X)$ (red line, left),  $p_{L,p}(X)$ (blue line, left), $p_{A,R,p}(X)$ (red line, right), and $p_{R,p}(X)$ (blue line, right)  in the Gumbel case. \label{fig:Gumbel}}
\end{figure}  

{\bf Gumbel $l$-type}. In order to obtain $p_{A,L,p}(X)$ and $p_{A,R,p}(X)$ in the whole this probability $l$-type it is enough to consider the simplest form of its c.d.fs. Therefore, now we assume that $X$ is a r.v. with c.d.f. $F_X(x) = \exp\{-\exp(-x)\}$, for all $x \in \mathbb{R}$. The last means that the quantile function is $F_X^\leftarrow(p) = (-\log(-\log(p))$. The theoretical asymmetric left and right $p$-fences are correspondingly
\begin{eqnarray*}
  L^A(X,p)  & = & -\frac{1}{p}\log(-\log(p)) + \frac{1-p}{p}\log(\log(2))\\
  R^A(X,p)  & = & -\frac{1}{p}\log(-\log(1-p)) + \frac{1-p}{p}\log(\log(2)).
  \end{eqnarray*}
After some algebraic computations the definitions 3 and 4 entail,
\begin{eqnarray*}
p_{A,L,p}(X) & = & \exp\left\{-\exp\left(\frac{1}{p}\log(-\log(p)) - \frac{1-p}{p}\log(\log(2))\right)\right\} = 2^{\left(-\left(-\log_2(p)\right)^{\frac{1}{p}}\right)}\\
p_{A,R,p}(X) & = & 1 - \exp\left\{-\exp\left(\frac{1}{p}\log(-\log(1-p)) - \frac{1-p}{p}\log(\log(2))\right)\right\} = 1 - 2^{\left(-(-\log_2(1 - p))^{\frac{1}{p}}\right)}.
  \end{eqnarray*}
  The plots of  $p_{A,L,p}(X)$, and $p_{L,p}(X)$, (\cite{jordanova2019general}) are presented in Figure \ref{fig:Gumbel}, left. Those of
   $p_{A,R,p}(X)$, and $p_{R,p}(X)$, could be seen in Figure \ref{fig:Gumbel}, right.
  
\begin{figure}[h]
\begin{minipage}[t]{0.5\linewidth}
\centerline{%
    \includegraphics[width=.86\textwidth]{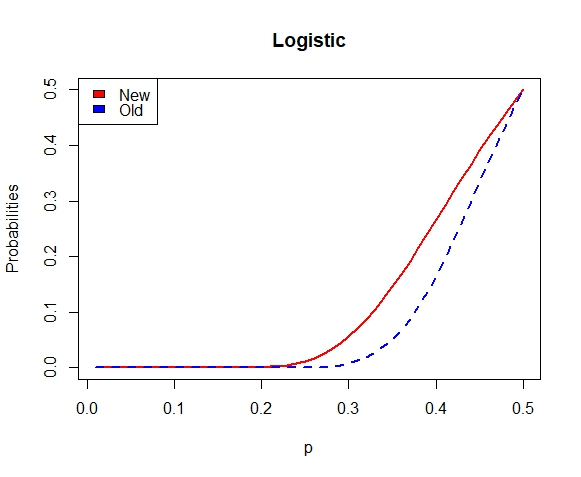}}
\end{minipage}
\begin{minipage}[t]{0.5\linewidth}
\centerline{%
    \includegraphics[width=.86\textwidth]{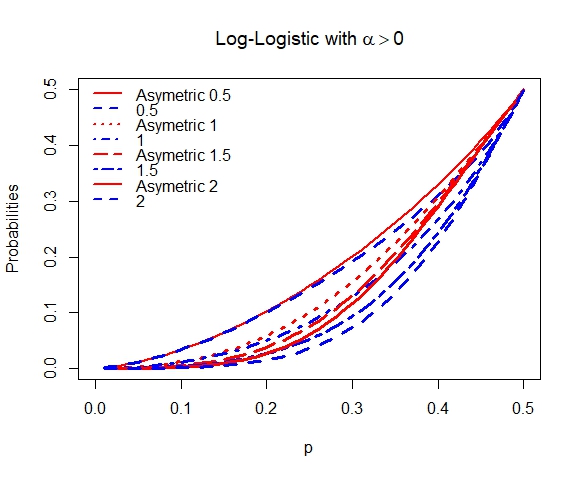}}
\end{minipage}
\caption{$p_{A,R,p}(X)$ (red lines) and $p_{R,p}(X)$ (blue lines) in the Logistic (left) and Log-Logistic (right) cases. \label{fig:LogisticAndLogLogistic}}
\end{figure}

{\bf Logistic $l$-type.}  Consider a r.v. $X$ with c.d.f. $F_X(x) = \frac{1}{1+\exp\{-x\}}$, $x \in \mathbb{R}$.  It is well-known that this distribution is symmetric, therefore $L^A(X,p) = -R^A(X,p)$, $p_{A,L,p}(X) = p_{A,R,p}(X)$. Its quantile function is $F_X^\leftarrow(p) = \log\left(\frac{p}{1-p}\right)$. By definitions 1 and 2 we have $L^A(X,p) = -R^A(X,p) = \frac{1}{p}\log\left(\frac{p}{1-p}\right)$. Thus, definitions 3 and 4 entail,
$$p_{A,L,p}(X) = p_{A,R,p}(X) = \frac{1}{1+\exp\left\{-\frac{1}{p}\log\left(\frac{p}{1-p}\right)\right\}} = \frac{1}{1+\left(\frac{1-p}{p}\right)^{\frac{1}{p}}}.$$
The plots of these probabilities, and $p_{L,p}(X) = p_{R,p}(X)$, as functions of $p$ are given in Figure \ref{fig:LogisticAndLogLogistic}, left. We observe that these distributions have very similar tail behaviour with the Gumbel distribution.

{\bf Log-Logistic $l$-type with parameter $\alpha > 0$.} Let $X$ be a r.v.  with c.d.f. $F_X(x) = \frac{1}{1+x^{-\alpha}}$, $x \geq 0$ and $F_X(x) = 1$, otherwise. This c.d.f. corresponds to a quantile function $F_X^\leftarrow(p) = \left(\frac{p}{1-p}\right)^{\frac{1}{\alpha}}$. Therefore,
\begin{eqnarray*}
  L^A(X,p)  & = & \frac{1}{p}\left(\frac{p}{1-p}\right)^{\frac{1}{\alpha}} - \frac{1-p}{p}. \quad L^A(X,p) \geq 0, \iff \alpha > \alpha_0, \quad {\text{ where }}\alpha_0 = \log_{1-p}(p) - 1 \geq 0.\\
   R^A(X,p) & = &  \frac{1}{p}\left(\frac{1-p}{p}\right)^{\frac{1}{\alpha}} - \frac{1-p}{p} \geq 0, \quad  \forall \alpha > 0,
   \end{eqnarray*}
$$p_{A,L,p}(X)  = \left\{\begin{array}{ccc}
                           \frac{1}{1+\left(\frac{1}{p}\left(\frac{p}{1-p}\right)^{\frac{1}{\alpha}} - \frac{1-p}{p}\right)^{-\alpha}} & , & \alpha > \alpha_0 \\
                           0 & , & 0 < \alpha \leq \alpha_0 
                         \end{array}\right., {\text{ and }} p_{A,R,p}(X)  =  \frac{\left(\frac{1}{p}\left(\frac{1-p}{p}\right)^{\frac{1}{\alpha}} - \frac{1-p}{p}\right)^{-\alpha}}{1+\left(\frac{1}{p}\left(\frac{1-p}{p}\right)^{\frac{1}{\alpha}} - \frac{1-p}{p}\right)^{-\alpha}}, \quad  \forall \alpha > 0. $$
The last probability, together with $p_{R,p}(X)$, are plotted in Figure \ref{fig:LogisticAndLogLogistic}, right.  They confirm that the tail behaviour of Log-Logistic distribution is very similar to the one of Pareto and Fr$\acute{e}$chet distributions.
\begin{figure}[h]
\begin{minipage}[t]{0.5\linewidth}
\centerline{\includegraphics[width=.86\textwidth]{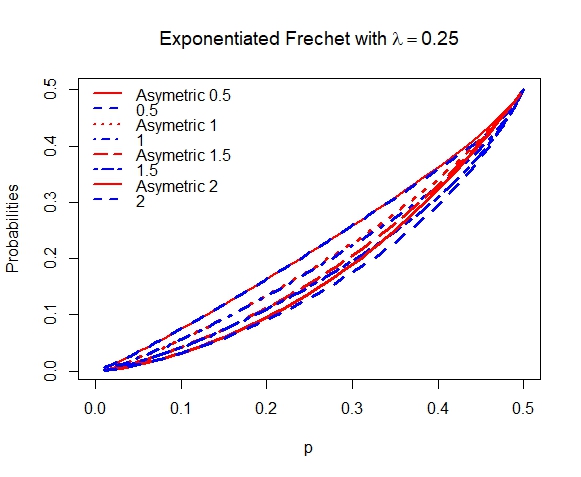}}
\end{minipage}
\begin{minipage}[t]{0.5\linewidth}
\centerline{\includegraphics[width=.86\textwidth]{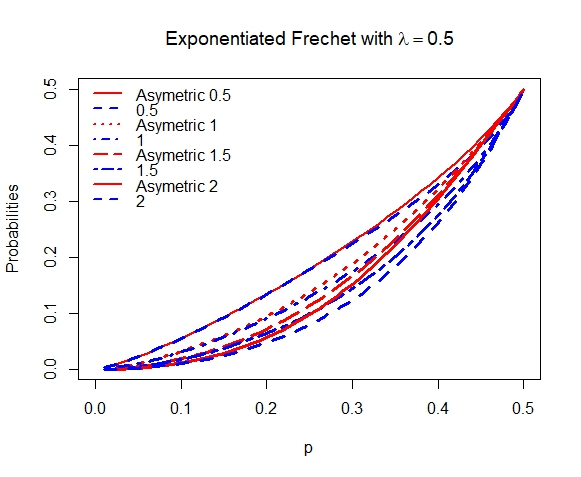}}
\end{minipage}
\caption{$p_{A,R,p}(X)$ (red lines) and $p_{R,p}(X)$ (blue lines) in the Exponentiated Fr$\acute{e}$chet case for different values of $\alpha$ and $\lambda$. \label{fig:ExponentiatedFrechet1}}
\end{figure}
\begin{figure}[h]
\begin{minipage}[t]{0.5\linewidth}
\centerline{\includegraphics[width=.86\textwidth]{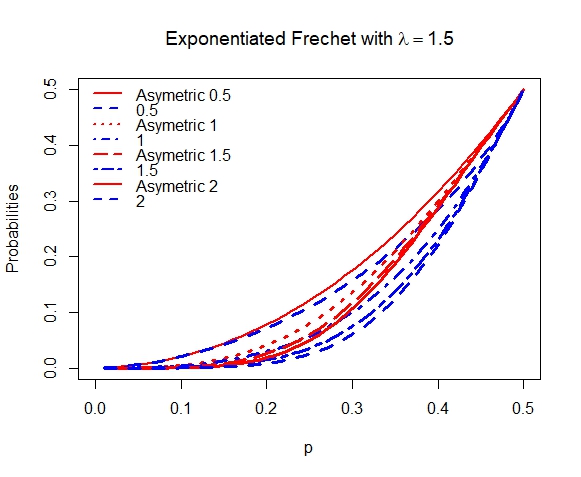}}
\end{minipage}
\begin{minipage}[t]{0.5\linewidth}
\centerline{\includegraphics[width=.86\textwidth]{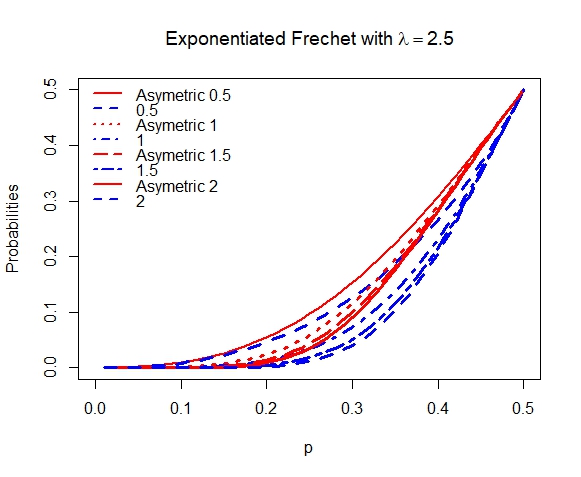}}
\end{minipage}
\caption{$p_{A,R,p}(X)$ (red lines) and $p_{R,p}(X)$ (blue lines) in the Exponentiated Fr$\acute{e}$chet case for different values of $\alpha$ and $\lambda$. \label{fig:ExponentiatedFrechet2}}
\end{figure}

{\bf Exponentiated Fr$\acute{e}$chet $l$-type with parameters $\alpha > 0$ and $\lambda > 0$.} This distribution was introduced in 2003 by Nadarajah and Kotz \cite{nadarajah2003exponentiated}. It is interesting because the heaviness of its tails depend of two parameters. The authors show that its tail is regularly varying function with parameter $-\alpha \lambda$. In order to see how this fact looks in the concepts for probabilities for asymmetric outside values, let us consider a r.v. $X$ with c.d.f. $F_X(x) = 1 - \left\{1 - \exp\{-x^{-\alpha}\}\right\}^\lambda$, $x \geq 0$ and $F_X(x) = 0$, otherwise. This c.d.f. corresponds to the quantile function $F_X^\leftarrow(p) = \{-\log[1-(1-p)^{\frac{1}{\lambda}}]\}^{-\frac{1}{\alpha}}$, which could be seen for example in \cite{jordanova2020probabilities}. In this case
\begin{eqnarray*}
  L^A(X,p)  & = & \frac{1}{p}\left\{-\log\left[1-(1-p)^{\frac{1}{\lambda}}\right]\right\}^{-\frac{1}{\alpha}} - \frac{1-p}{p}\left\{-\log\left[1-2^{-\frac{1}{\lambda}}\right]\right\}^{-\frac{1}{\alpha}}\\
  R^A(X,p)  & = & \frac{1}{p}\left\{-\log\left[1-p^{\frac{1}{\lambda}}\right]\right\}^{-\frac{1}{\alpha}} - \frac{1-p}{p}\left\{-\log\left[1-2^{-\frac{1}{\lambda}}\right]\right\}^{-\frac{1}{\alpha}}.
  \end{eqnarray*}
\begin{figure}[h]
\begin{minipage}[t]{0.5\linewidth}
\centerline{\includegraphics[width=.86\textwidth]{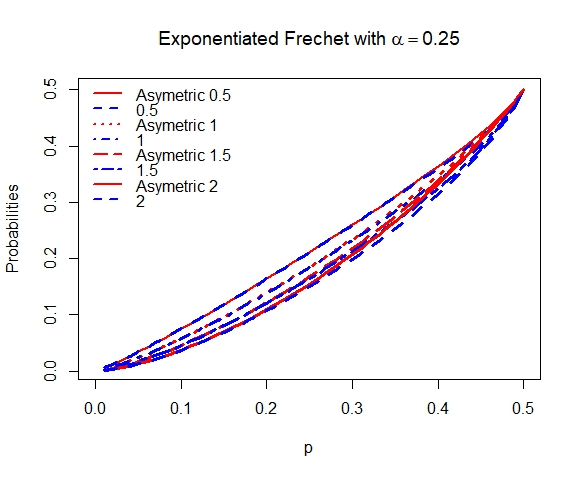}}
\end{minipage}
\begin{minipage}[t]{0.5\linewidth}
\centerline{\includegraphics[width=.86\textwidth]{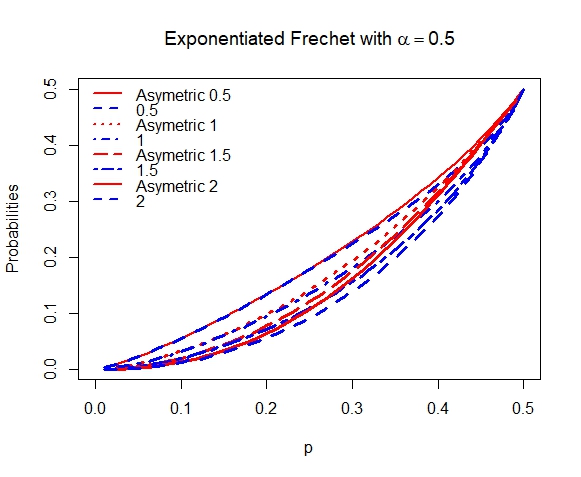}}
\end{minipage}
\caption{$p_{A,R,p}(X)$ (red lines) and $p_{R,p}(X)$ (blue lines) in the Exponentiated Fr$\acute{e}$chet case for different values of $\alpha$ and $\lambda$. \label{fig:ExponentiatedFrechet3}}
\end{figure}
\begin{figure}[h]
\begin{minipage}[t]{0.5\linewidth}
\centerline{\includegraphics[width=.86\textwidth]{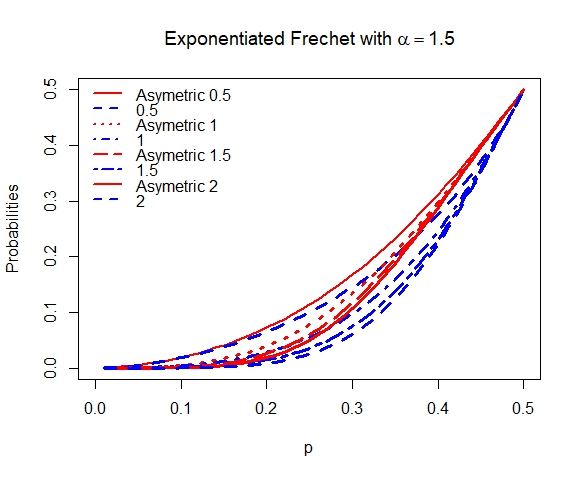}}
\end{minipage}
\begin{minipage}[t]{0.5\linewidth}
\centerline{\includegraphics[width=.86\textwidth]{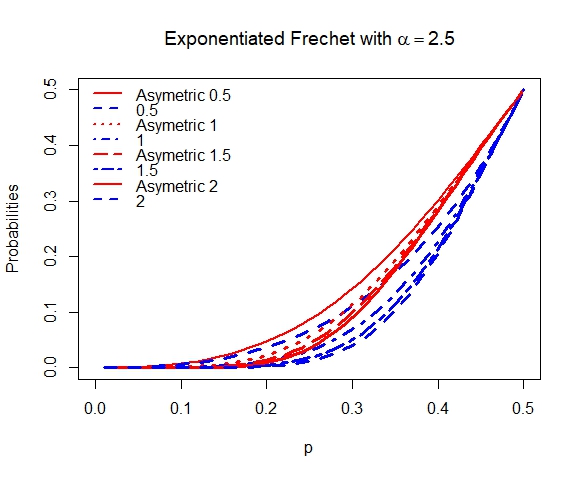}}
\end{minipage}
\caption{$p_{A,R,p}(X)$ (red lines) and $p_{R,p}(X)$ (blue lines) in the Exponentiated Fr$\acute{e}$chet case for different values of $\alpha$ and $\lambda$. \label{fig:ExponentiatedFrechet4}}
\end{figure}

In order to use the c.d.f. of $X$, we need first to solve the inequality $L^A(X,p) \geq 0$. It is equivalent to the inequality 
$1-(1-p)^{\frac{1}{\lambda}} \geq \left(1-2^{-\frac{1}{\lambda}}\right)^{(1-p)^{-\alpha}}.$
Let us denote by $p_0$ the solution of the equation $1-(1-p)^{\frac{1}{\lambda}} = \left(1-2^{-\frac{1}{\lambda}}\right)^{(1-p)^{-\alpha}}.$
Then, $p_{A,L,p}(X) = 0$ when $p \in (0, p_0]$, and 
$$p_{A,L,p}(X)  =  1 - \left\{1 - \exp\left\{-\left(\frac{1}{p}\left\{-\log\left[1-(1-p)^{\frac{1}{\lambda}}\right]\right\}^{-\frac{1}{\alpha}} - \frac{1-p}{p}\left\{-\log\left[1-2^{-\frac{1}{\lambda}}\right]\right\}^{-\frac{1}{\alpha}}\right)^{-\alpha}\right\}\right\}^\lambda, \quad p \in (p_0, 0.5].$$

It is easy to check that $R^A(X,p) \geq 0$ in the whole parametric space. Therefore, by using definitions 3 and 4 one obtain that for any $\lambda > 0$, $\alpha > 0$ and $p \in (0; 0.5]$, 
$$p_{A,R,p}(X) =  \left\{1 - \exp\left\{-\left(\frac{1}{p}\left\{-\log\left[1-p^{\frac{1}{\lambda}}\right]\right\}^{-\frac{1}{\alpha}} - \frac{1-p}{p}\left\{-\log\left[1-2^{-\frac{1}{\lambda}}\right]\right\}^{-\frac{1}{\alpha}}\right)^{-\alpha}\right\}\right\}^\lambda.$$
 The dependencies of these probabilities on $p$, for different values of $\lambda$ and $\alpha$ are depicted in Figures \ref{fig:ExponentiatedFrechet1}-\ref{fig:ExponentiatedFrechet4}. We observe that the smaller the value of the products of $\alpha$ and $\lambda$, the higher the probabilities for asymmetric $p$-outside values are.
 
\begin{figure}[h]
 \begin{minipage}[t]{0.5\linewidth}
\centerline{\includegraphics[width=.86\textwidth]{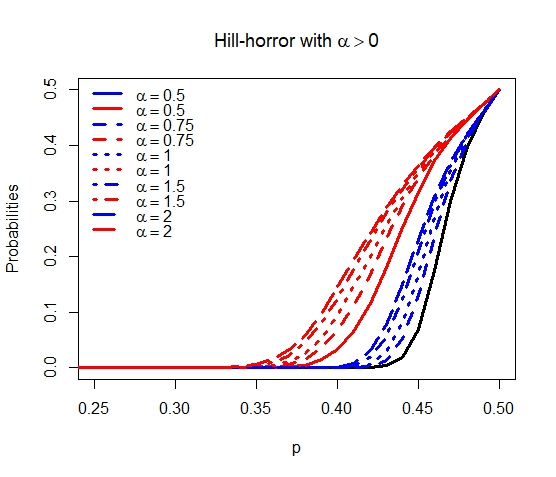}}
\end{minipage}
\begin{minipage}[t]{0.5\linewidth}
\centerline{\includegraphics[width=.86\textwidth]{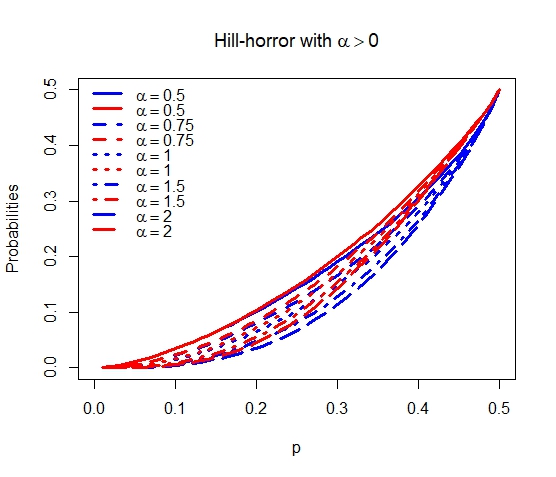}}
\end{minipage}
\caption{Dependence of $p_{L,p}(X)$ (left, blue), $p_{A,L,p}(X)$ (left, red), $p_{R,p}(X)$ (right, blue), and  $p_{A,R,p}(X)$ (right, red) on $p$ and $\alpha$ in the case when the r.v. $X$ is Hill-horror distributed \label{fig:HillH}}
\end{figure}

 {\bf Hill-horror $l$-type with parameter $\alpha > 0$.} The definition of this distribution could be seen e.g. in \cite{EKM}. It is based on the quantile function $F^\leftarrow(p) = (1 - p)^{-1/\alpha}(-\log\,(1 - p))$, and it is an example of a probability law when the probabilities for asymmetric $p$-outside values can be determined, but the explicit form of the c.d.f. is not known. Analogously to Example 2.18, in p. 51, \cite{jordanova2020probabilities}, in order to compute them, we use possibilities for simulations of R statistical software (R Core Team, 2024) \cite{r2013r}. We estimate the c.d.f. of the r.v. $X$, $p_{A,L,p}(X)$ and $p_{A,R,p}(X)$. The plots of the last two functions of $p$, for some  different values of $\alpha$, are presented in Figure \ref{fig:HillH}. 
 
In order to make these plots first we have simulated $1000$ samples of $1000000$ independent observations on $X$. Then, we
have estimated the c.d.f. of $X$, and finally we have determined the plotted values. As far as for any fixed $\alpha$, Hill-horror, Pareto, Fr$\acute{e}$chet, and Log-Logistic distributions belong to one and the same class of heavy-tailed distributions with $RV_\alpha$ right tails (See e.g. Embrechts et al. \cite{EKM}), the probabilities
for asymmetric $p$-outside values are very similar in these plots.

\section{A new classification of linear probability types with respect to their tail behaviour}
It is well-known that in cases when it exists, the tail index governs the tail behaviour of the corresponding probability c.d.f. independently on the location and the scale parameters. The same is true for probabilities for asymmetric right $p$-outside values. However, these probabilities  allow us to insert a larger class of distributions in this partial ordering than the classification of heavy-tailed distributions based on the extremal index. Here, we fix $p = 0.25$, and show that the probabilities for asymmetric right $0.25$-outside values can be used to classify distributions with respect to their tail behaviour.  To simplify our considerations we discuss only absolutely continuous distributions with infinite supports. 

\begin{figure}[h]
\centerline{\includegraphics[scale=.6]{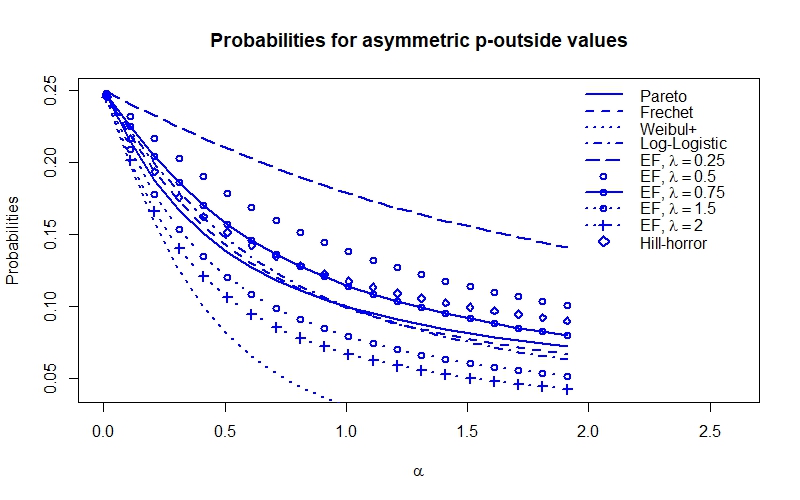}}
\caption{Dependence of $p_{A,R,p}(X)$ on $\alpha$ \label{fig:DependenceOnAlpha}}
\end{figure}

By using the results in previous section we obtain that for all  distributions in the Exponential linear probability type, the probability for asymmetric $0.25$-outside values is $p_{A,R,p}(X) \approx 0.0313$. For the Gumbel linear type it is $\approx 0.0204$. For the Logistic case it is $\approx 0.0122$. The dependence of probabilities for asymmetric $0.25$-outside values on the parameter $\alpha = \frac{1}{\xi}$ within the set of the another considered distributions in this work can be seen in Figure \ref{fig:DependenceOnAlpha}.

The classification based on probabilities for asymmetric $0.25$-outside values confirms the corresponding well-known results based on the concepts for light and heavy-tailed distributions, and the extremal index. Its advantage is that it is more detailed, and therefore, more useful in practice. It outperforms the popular concept for kurtosis as far as in many of the considered cases the kurtosis does not exist.

\section{Two distribution sensitive estimators of the parameters which govern the tail behaviour. Simulation study.}

Here we assume that we have a sample of i.i.d observations, there is at least one observation, which is bigger than the empirical asymmetric right $p$-fence, and $R_n^A(p) > 1$. Thus, we are working given $X_{n:n} > R_n^A(p) > 1$. Depending on the values of the plug-in estimators of $p_{A,R,p}(X)$ and $p_{A,L,p}(X)$ we chose the most appropriate probability types for modelling the observed r.v. Then, analogously to the method of moments and \cite{jordanova2017measuring},  by using the expected equality between the estimated value of $p_{A,R,p}(X)$, and the theoretical formula for $p_{A,R,p}(X)$, we express the unknown parameter and obtain the formula for the estimator. In order to simplify the computations empirical asymmetric right $p$-fence can be used, if needed. Finally, we can use some goodness of fit test in order to chose the best model.
It is clear that for application of this algorithm we need the explicit form of $p_{A,R,p}(X)$. 

We consider two cases, and obtain the following distribution sensitive estimators. 

{\bf Case 1}. If the observed r.v. $X$ is $Pareto(\alpha)$, $\alpha > 0$ distributed we apply
\begin{equation}\label{ParetoEst}
\widehat{\alpha}_{Par,n} (p)=  -\frac{\log\, \left(\widehat{p}_{A,R}(p)\right)}{\log\,\left(\widehat{R}_n^A(p)\right)}.
\end{equation}

{\bf Case 2}. For $X \in Fr\acute{e}chet(\alpha)$, $\alpha > 0$ the corresponding estimator is
\begin{equation}\label{FrechetEst}
\widehat{\alpha}_{Fr, n}(p) =  -\frac{\log\left(-\log\, \left(1-\widehat{p}_{A,R}(p)\right)\right)}{\log\,\left(\widehat{R}_n^A(p)\right)}.
\end{equation}

The best value of $p$ seems to be around the middle of the interval $(0; 0.5]$, because in that case the sensitivity of the $p_{A,R,p}(X)$ with respect to different probability laws is the best. Therefore, $p$ have to be close to $0.25$. This section finishes with $p = 0.25$. First, we simulated $n$ independent observations on $Pareto(\alpha)$, and separately $n$ independent observations on   $Frechet(\alpha)$ distribution, for different values of  $\alpha \in (0; 2]$, and for different but fixed $n = 30, 10^2, 10^3, 10^4$. Then, we have repeated this experiments $m = 10^4$ times. In this way, we have calculated $10^4$ values of $\hat{\alpha}_{Par,n}$, and $10^4$ values of $\hat{\alpha}_{Frech, n}$, their means and standard deviations. The results are given in Table~\ref{tab:estimators}.

Analogously to \cite{jordanova2017measuring} we observe that if the sample size is $n > 1000$, then the better estimator in any particular case is the one that takes into account the probabilistic type of the observed r.v. Therefore, when we are using these estimators, the choice of the most appropriate distribution is the an important previous step before the estimation of the parameter which governs the tail behaviour.

\begin{table}
\centering
\caption{Empirical results.   \label{tab:estimators}}

\begin{tabular}{|c|c|c|c|c|c|c|c|}
  \hline
  \multicolumn{2}{|c|}{Distribution}  & \multirow{2}{*}{$n$}  &\multicolumn{2}{c|}{$\hat{\alpha}_{Par,n}$}   & \multicolumn{2}{c|}{$\hat{\alpha}_{Frech, n}$}    & \multirow{2}{*}{Better}\\
  \cline{4-7}
 \multicolumn{2}{|c|}{of $X$.}   &                           & Mean & St. Dev. & Mean & St. Dev. & estimator\\ \hline
\multirow{12}{*}{$Pareto(\alpha)$} & \multirow{4}{*}{$\alpha = 0.5$}  & $30$ & 0.5456 & 0.1339&  {\bf 0.5286} & 0.1356 & $\widehat{\alpha}_{Fr,n}$\\
                                   &                                  & $100$ &0.5136& 0.0659 & {\bf 0.4953} & 0.0669 & $\widehat{\alpha}_{Fr,n}$\\
                                   &                                  & $1000$ & {\bf 0.5011} & 0.0199 &  0.4825 & 0.0203 &  $\widehat{\alpha}_{Par,n}$\\
                                   &                                  & $10000$ & {\bf 0.5002} & 0.0063 &  0.4815 & 0.0064 & $\widehat{\alpha}_{Par,n}$\\
  \cline{2-8}
                                   & \multirow{4}{*}{$\alpha = 1$}  & $30$ & 1.0875 & 0.2639&  {\bf 1.0654} & 0.2671 & $\widehat{\alpha}_{Fr,n}$\\
                                   &        &                         $100$ &1.0297& 0.1407 & {\bf 1.0071} & 0.1433 & $\widehat{\alpha}_{Fr,n}$\\
                                   &        &                        $1000$ & {\bf 1.0022} & 0.0413 & 0.9794 & 0.0421 & $\widehat{\alpha}_{Par,n}$\\
                                   &        &                       $10000$ & {\bf 1.0001} & 0.0131 &  0.9774& 0.0134 &$\widehat{\alpha}_{Par,n}$\\
  \cline{2-8}
                                   & \multirow{4}{*}{$\alpha = 2$}  & $30$ & 2.1178 & 0.5018&  {\bf 2.0862} & 0.5056 & $\widehat{\alpha}_{Fr,n}$\\
                                   &        &                         $100$ &2.0686& 0.3015 & {\bf 2.0430} & 0.3065 & $\widehat{\alpha}_{Fr,n}$\\
                                   &        &                        $1000$ & {\bf 2.0073} & 0.0869 &  1.9797 & 0.0884 & $\widehat{\alpha}_{Par,n}$\\
                                   &        &                       $10000$ & {\bf 2.0005} & 0.0272 &  1.9729 & 0.0277 &$\widehat{\alpha}_{Par,n}$\\
   \hline
 \multirow{12}{*}{$Frechet(\alpha)$} & \multirow{4}{*}{$\alpha = 0.5$}  &  $30$ & 0.5753 & 0.1558&  {\bf 0.5566} & 0.1560 & $\widehat{\alpha}_{Fr,n}$\\
                                  &        &                         $100$ &0.5366& 0.0749 & {\bf 0.5166} & 0.0753 & $\widehat{\alpha}_{Fr,n}$\\
                                   &        &                        $1000$ & 0.5216 & 0.0223 & {\bf 0.5012} & 0.0224 & $\widehat{\alpha}_{Fr,n}$\\
                                   &        &                       $10000$ &  0.5205 & 0.0007 &  {\bf 0.5002} & 0.0071 &$\widehat{\alpha}_{Fr,n}$\\
   \cline{2-8}
  & \multirow{4}{*}{$\alpha = 1$} &                                   $30$ & 1.1144 & 0.2870&  {\bf 1.0916} & 0.2892 & $\widehat{\alpha}_{Fr,n}$\\
                                  &        &                         $100$ &1.0561& 0.1476 & {\bf 1.0334} & 0.1499 & $\widehat{\alpha}_{Fr,n}$\\
                                   &        &                        $1000$ & 1.0263 & 0.0439 & {\bf 1.0034} & 0.0446 & $\widehat{\alpha}_{Fr,n}$\\
                                   &        &                       $10000$ &  1.0232 & 0.0014 &  {\bf 1.0003} & 0.0142 &$\widehat{\alpha}_{Fr,n}$\\
   \cline{2-8}
   & \multirow{4}{*}{$\alpha = 2$} &                                  $30$ & 2.1358 & 0.5200&  {\bf 2.1061} & 0.5223 & $\widehat{\alpha}_{Fr,n}$\\
                                  &        &                         $100$ &2.1017& 0.3188 & {\bf 2.0765} & 0.3235 & $\widehat{\alpha}_{Fr,n}$\\
                                   &        &                        $1000$ & 2.0321 & 0.0923 & {\bf 2.0073} & 0.0938 & $\widehat{\alpha}_{Fr,n}$\\
                                   &        &                       $10000$ &  2.0256 & 0.0289 & {\bf 2.0008} & 0.0294 &$\widehat{\alpha}_{Fr,n}$\\
   \hline
\end{tabular}
\end{table}

\section{Conclusive remarks and open problems}
This work improves the results in \cite{jordanova2017measuring,jordanova2018tails,jordanova2019general,jordanova2019probabilities}. It states that probabilities of the events an observation to be an asymmetric $p$-outside value can be used as universal characteristics of the heaviness of the tail of the considered c.d.f. They  always exist, and are invariant within all linear distributional type. They are applicable in a wider class of distributions than the class of c.d.fs. with regularly varying tails. They outperform the kurtosis, because in the case of heavy-tails the kurtosis does not exist. They can be used for data visualisation, as far as, their theoretical values in many particular cases have explicit forms and the corresponding empirical functions can be plotted and compared. This classification refines the well-known classification based on the extremal index in cases when it exists. For distributions with heavier tails the estimators give better results. Their relatively fast rate of convergence allows one to apply them also for relatively small samples.

It is interesting to analyse the role of the constants $p_L(X) := \inf\{p > 0: p_{A,L,p}(X) > 0 \}$ and $p_R(X) := \inf\{p > 0: p_{A,R,p}(X) > 0 \}$ in the tail behaviour of the c.d.f. Our experiments and plots of $p_{A,L,p}(X)$ and $p_{A,R,p}(X)$ w.r.t. $p$ show that they give some inside of the answer of the question "Has the observed distribution heavy-tail?", and in this way contribute to understanding of the tail behaviour.  For example, if for all $p \in (0; 0,5]$, $p_{A,L,p}(X) >  0$ ($p_{A,R,p}(X) > 0$), then the observed c.d.f. has heavy left (right) tail. Our hypothesis is that this is a necessary and sufficient condition.

And last but not least, an important open question is to determine the most appropriate value of $p$ in order to obtain the best estimators for the parameters which govern the tail behaviour.

\section{ACKNOWLEDGMENTS}
The author was partially supported by the Project RD-08-.../.......202.... from the Scientific Research Fund in
Konstantin Preslavsky University of Shumen, Bulgaria.


\end{document}